# Arterial Blood Pressure Feature Estimation Using Photoplethysmography

Armin Soltan Zadi[1], Raichel Alex[1], Rong Zhang[2], Donald E. Watenpaugh[3], Khosrow Behbehani[1]

[1]The University of Texas at Arlington, Arlington, Texas
[2]The University of Texas Southwestern Medical Center, Dallas, Texas
[3]Sleep Consultants Inc., Fort Worth, TX

Corresponding Author: Armin Soltan Zadi
Email: armin.soltanzadi@uta.edu
Phone: (214)971-0901

*Abstract*— Continuous and noninvasive monitoring of blood pressure has numerous clinical and fitness applications. Current methods of continuous measurement of blood pressure are either invasive and/or require expensive equipment. Therefore, we investigated a new method for the continuous estimation of two main features of blood pressure waveform: systolic and diastolic pressures. The estimates were obtained from a photoplethysmography signal as input to the fifth order autoregressive moving average models. The performance of the method was evaluated using beat-to-beat full-wave blood pressure measurements from 15 young subjects, with no known cardiovascular disorder, in supine position as they breathed normally and also while they performed a breath-hold maneuver. The level of error in the estimates, as measured by the root mean square of the model residuals, was less than 5 mmHg during normal breathing and less than 8 mmHg during the breath-hold maneuver. The mean of model residuals both during normal breathing and breath-hold maneuvers was considered to be less than 3.2 mmHg. The dependency of the accuracy of the estimates on the subject data was assessed by comparing the modeling errors for the 15 subjects. Less than 1% of the models showed significant differences ($p < 0.05$) from the other models, which indicates a high level of consistency among the models.

## I. INTRODUCTION

HYPERTENSION is one of the major risk factors contributing to cardiovascular morbidity and mortality of patients [1] [2]. Noninvasive, continuous blood pressure (BP) monitoring has been shown to be a superior detector of mortality than clinical measurements [3]. Direct, continuous blood pressure measurements using cuff-based methods are robust and reliable, but they work discontinuously, requiring a few minutes for each measurement [4]. Most of the traditional methods for indirect continuous blood pressure monitoring work based on auscultation and/or oscillation, yet perform intermittently [5]. Most widely used methods are based on vascular unloading proposed by Penaz 40 years ago [6]. A Finapres machine works by applying varying pressure via a finger cuff and using a fast servomotor to react to changes in blood volume measured by a photoplethysmograph (PPG) [7]. However, despite several improvements by others [8] [9] [10] [11] it still suffers from frequent interruptions for calibration, which can be a major drawback, especially if it happens during transient changes in BP. Moreover, using a finger cuff can be cumbersome and disruptive during certain conditions such as exercise or sleep studies. On the other hand, tonometry methods do not suffer from the disadvantages of using in line cuffs, as they use piezoelectric crystals to sense oscillations [12] [13]; however, manual applanation has its own difficulties. Both tonometry and auscultatory methods are subject to fixation and highly sensitive to motion artifacts.

Recent studies have highlighted the benefits of using PPG signals and their features to estimate BP changes indirectly. The most widely used feature is pulse transit time (PTT), extracted from PPG and electrocardiogram (ECG). The correlation between systolic and diastolic BP measurements and PTT has been shown to be more than 0.8 in experimental studies [14] [15] [16]. While a relatively high correlation is observed between the estimates of BP from PTT, the computation of PTT requires the availability of an ECG signal, which necessitates having added instrumentation compared to the methods that use PPG alone. Aside from additional costs for using another device, synchronization is a major issue in using these signals in real time. Further, some researchers have reported that PTT may not always be a reliable marker of beat-to-beat blood pressure due to the dynamic nature of human muscles and hydrostatic changes [17]. Consequently, it is desirable to explore other methods of using PPG signal alone to estimate BP.

Several studies have reported approaches to direct estimation of BP from PPG. A simple yet efficient method is to utilize a regression model to relate PPG amplitude to BP, which has been shown to be a suitable method [18]. However, while this method shows promising results for systolic BP estimation, it does not perform as well for diastolic BP estimation. Further, since regression has no memory it cannot model the delay between PPG and BP. Therefore, there may remain some potential for obtaining estimates of BP with higher accuracy by exploring the benefits of integrating more information from PPG into the estimation of BP. An alternative approach has been to use the Fourier domain analysis [19]. This method, while quite suitable for modeling a stationary system, is computationally complex. Further, neither BP nor PPG are stationary. As a result, the applicability of this method may be limited to certain conditions. In particular, if one desires to estimate BP from PPG when perturbations such as hypoxia are present, the needed stationarity assumption may not hold. Another study proposed some novel features to be used for BP estimation and reported reasonable accuracy [20]. They used the width of a half pulse amplitude, width of a two-third pulse amplitude, systolic upstroke time, and diastolic time of pulse in their research and achieved a mean difference of errors of 0.21 mmHg for systolic BP (SBP) and 0.02 mmHg for diastolic BP (DBP), with corresponding standard deviations of 7.32 mmHg for SBP and 4.39 mmHg for DBP.

Driven by the advances in miniaturization, a study has reported the use of multiple PPG sensors in wearable devices in an attempt to estimate BP from pulse wave velocity (PWV) [21]. PWV is defined as the distance between sensors, usually worn at the wrist and finger, divided by pulse transit time calculated from pulse onset from two sensors. Validation has been performed on only one subject, it showed high correlation of PWV and arterial BP but needs further investigation. Other wearable devices can estimate blood pressure, but they require an ECG and PPG to extract PTT as mentioned before. One study using regression to estimate both SBP and DBP reported root mean squared error (rMSE) to be between 7.83 to 9.37 mmHg for SBP and between 5.77 to 6.90 mmHg for DBP across 11 subjects [22]. Another study investigated continuous estimation of heart rate and SBP from PTT and reported a rMSE value of 4.71 mmHg, as well as mean and standard deviation of error of 1.63 ± 4.44 mmHg, with mean absolute error of 3.68 mmHg [23]. A more complex method, proposed by Datta *et al.* [24], is based on using the Windkessel model and tries to find its parameters from the PPG features. The authors claim they achieved errors in the range of 10% of a commercially available digital BP monitoring device.

In contrast to these other methods, autoregressive moving average (ARMA) modeling provides a means of mathematically relating the dynamic relationship between the input and output by incorporating possible influences of the present and previous values of the input as well as previous values of the output on the present value of the output. Additionally, ARMA modeling can take into account pure time delays that may be present in the process of relating input to output, a rather common phenomenon in physiological systems [25]. Furthermore, ARMA models allow for relating multiple inputs to either a single output or to multiple outputs [26]. This provides the user the capability of investigating the benefits of integrating multiple inputs when estimating the desired output. Researchers have investigated ARMA models in many other physiological applications such as cardiovascular and respiratory research [27] [28] [29] [30]. For this reason, the ARMA modeling method is a promising approach for estimating BP from PPG.

This study will use peaks and troughs of a PPG signal to estimate the key features from continuous blood pressure, namely SBP, DBP and mean arterial BP (MAP) using ARMA models. Sufficiency of the SBP, DBP, and computed MAP as a predictor of mortality has been shown in many studies [31] [32] [33] [34]. As part of this study, we investigated the modeling of BP from PPG when hypoxia is present and acts as a perturbation to the blood pressure physiological control system. The motivation was to accurately estimate and track the changes in BP by using a photoplethysmograph probe, which is easier to apply and less costly. Description of the PPG principles, proposed ARMA modeling and experimental assessment of the proposed models will be discussed in detail in the following sections.

## II. METHODS

In this study, we used peaks and troughs of the PPG waveform to model SBP and DBP, correspondingly, using autoregressive moving average (ARMA) models. Using the estimated values of systolic and diastolic BP, we estimated the MAP. We used a Finapres blood pressure NOVA monitor (Finapres Medical Systems, Enschede, Netherlands) for BP measurement as it has been validated and widely used before [35]. The photoplethysmography signal was acquired using a Nellcor OxiMax N-600x monitor (Medtronic, Minneapolis, USA). Both BP and PPG signals were acquired at 100 samples per second. The signals were clean and no application of noise reduction was necessary.

### A. Features of BP signal

For this study, we opted to use SBP, DBP, and MAP since these have been proven by wide use in clinical practice as robust indicators of blood pressure health. For this purpose, peaks of PPG signal were used to model the SBP and troughs of PPG signal were used for DBP modeling. Measured SBP points were calculated using the *findpeaks* algorithm of MATLAB (MathWorks Inc., Natick, MA, USA) with *MinPeakProminence* of 15 mmHg, *MinPeakDistance* of 20 samples, and *MinPeakHeight* of 15 mmHg. Then, minimum value between each two consecutive systolic values was considered to be the diastolic point of BP. Using the estimated SBP and DBP, we computed the estimate of MAP using the following:

$$MAP = \frac{2 \times DBP + SBP}{3} \quad (1)$$

### B. ARMA modeling

The ARMA model is a system identification method that can provide mathematical models of dynamic systems. The method allows for modeling of the dynamics of the system as well as any pure time delay [25]. For the purposes of this study, we applied a single input (i.e. PPG feature) and single output (i.e. BP feature). A single-input-single-output, time-invariant, and causal ARMA model can be represented as a dynamic difference equation involving present and past values of the input and output as described in the following equation (2):

$$y(m) + a_1 y(m-1) + \ldots + a_{n_a} y(m-n_a) = b_1 u(m-n_k) + \ldots + b_{n_b} u(m-n_b-n_k+1) + e(m) \quad (2)$$

Where, m is the sample number, y is the output, u is the input, and e stands for error, $a_i$ for $i = 1,2 \ldots, n_a$ and $b_j$ for $j = 1,2, \ldots, n_b$ are the model parameters that need to be computed, and $n_a$ and $n_b$ signify the orders of ARMA model while $n_k$ is the number of pure-time delay samples. A major decision in developing an effective ARMA model is the selection of the order of the model and any pure-time delay that may be involved. Investigators developing ARMA models have proposed guidelines for selecting these critical model parameters [36] [37] [38]. In this study, we used the principle of parsimony and model adequacy to find the least mean square error in a certain range of orders. Orders were set from 1 to 5, and delay ranged from 0 to 5 for our study. Then, the ARMA model parameters were estimated in MATLAB using a least square method that uses QR-factorization to minimize the model error for specific orders. Based on the selected values of $n_a$ and $n_b$ and $n_k$, all models of various orders were generated and compared using their residual MSE. The lowest MSE was selected as the best model generated in the specified range of $n_a$ and $n_b$ and $n_k$. We also used Akaike's information criterion (AIC) [36] [39] to find the best model, but it does not necessarily give the lowest MSE. The AIC finds the best model based on both parsimony and closeness of fit. In our study, we limited the model orders to be less than or equal to 5. This upper limit for the model order was selected based on examining MSE for models of various orders derived from a few randomly selected experimental settings (i.e. BH's (Breath hold) and NB's (Normal breathing)), which showed that model orders higher than 5 did not produce significant improvement in MSE.

Applying the ARMA model (Eq. 2) with the peaks of the PPG signal as input and SBP as output, we obtained the model parameters for estimating SBP (i.e. $\widehat{SBP}$) from a PPG signals. Similarly, by using troughs of PPG as input and DBP as output, we obtained the model parameters for estimating DBP (i.e. $\widehat{DBP}$) from PPG signals. Then, for both measured MAP and estimated MAP (i.e. $\widehat{MAP}$), we applied (1) by using respective measured (i.e. SBP and DBP) and estimated values (i.e. $\widehat{SBP}$ and $\widehat{DBP}$). Of course, estimated MAP reflected the measured MAP, which can be computed using an equation (Fig. 1) by substituting the measured values of SBP and DBP obtained from the blood pressure monitor (i.e. Finapres). Since the application of the ARMA model requires equidistant sampling of both the input and output data, we used cubic spline interpolation to interpolate DBP, SBP, and MAP values at the same sampling rate of the BP and PPG signal, i.e. 100 Hz. ARMA models for each of the five BH and six NB intervals were calculated separately for both systolic and diastolic BP in all 15 subjects. In other words, we obtained 11 SBP models and 11 DBP models for each subject.

*C. Photoplethysmography*

Photoplethysmography is a low-cost noninvasive method that uses optical sensors with two different wavelengths to detect the blood volume change in microvessels, typically from a finger. The PPG waveform shows a pulsatile waveform resulting from beat to beat cardiac cycle and a slow frequency baseline, which varies due to sympathetic nervous system response, respiration, and thermoregulation [40] [41]. Both blood oxygenation and blood volume changes can affect the pulsatile PPG waveform while the most dominant factor is shown to be blood volume change in peripheral vasculature rather than blood oxygenation [40] [47]. Consequently, it is reasonable to expect that PPG signal will contain significant information about blood pressure.

*D. Experimental setup*

The protocol and written subject consent form for testing subjects was approved by our Human Subject Institutional Review Board. Fifteen young subjects (eight males, seven females, aged 28.9 ± 5.0 years, BMI 24.1 ± 4.8 kg/m$^2$) with no known medical conditions volunteered for this study and signed the consent form. None of the subjects showed hypertension based on evaluation of their baseline data and comparison with the published values by the Centers for Disease Control and Prevention (CDC, an official U.S. government public health monitoring agency). The average of their systolic and diastolic values turned out to be 126.8 and 74.8 mmHg, respectively. The subjects were asked to avoid any caffeine intake for six hours before the experiments. The subjects were tested in supine position, performing a sequence of breath-holding maneuvers. The sequence of the maneuvers is shown in Fig. 1. At the start, while lying down on a bed, each subject was asked to breathe normally for 60 s to obtain baseline data. Afterward each subject performed a series of five breath-hold maneuvers to induce dynamic changes in BP. During each breath-hold (BH), subjects were instructed to hold their breath for as long as they could. This resulted in varying durations for each breath-hold, dependent on the ability of the subject to prolong it. Inter-breath-hold intervals were fixed at 90 seconds to provide adequate recovery time between consecutive breath-holds. At the conclusion of the fifth breath-hold maneuver, the subject remained in position and resumed breathing normally for 60 seconds. Data was collected for the entire duration, from the initial baseline through the final ending NB period (Fig. 1).

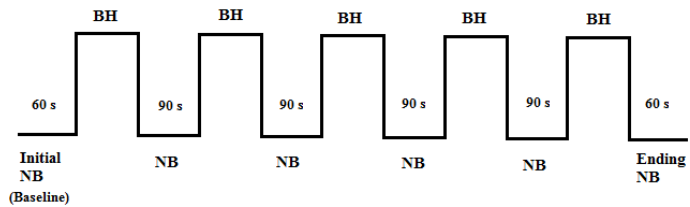

Figure 1- Timing diagram of the breath-hold (BH) maneuvers and normal breathing (NB) intervals

The breath-hold maneuver protocol was selected as it simulates sleep apnea and elicits the physiological responses that sleep apnea events do, such as drop in oxygen saturation and rise in carbon dioxide. Indeed, these changes stimulate the chemoreceptors that in turn trigger the sympathetic nervous system to increase arterial blood pressure. A typical BP and PPG signal are shown in Fig. 2, which were acquired from Finapres and OxiMax N-600x, respectively, as discussed before. The green line shows the breath-holds intervals whenever it is at the higher level. At its low values, the subject is in normal breathing condition. As shown, the blood pressure follows a rising trend during BHs, and so does the PPG amplitude. It is noted that irrespective of the duration of each breath-hold, the blood pressure and the level of PPG tended to rise and reach the highest level when BH was ended or shortly afterwards. The rise in PPG may be attributable to a greater volume of blood reaching to the peripheral vasculature as the blood pressure rises.

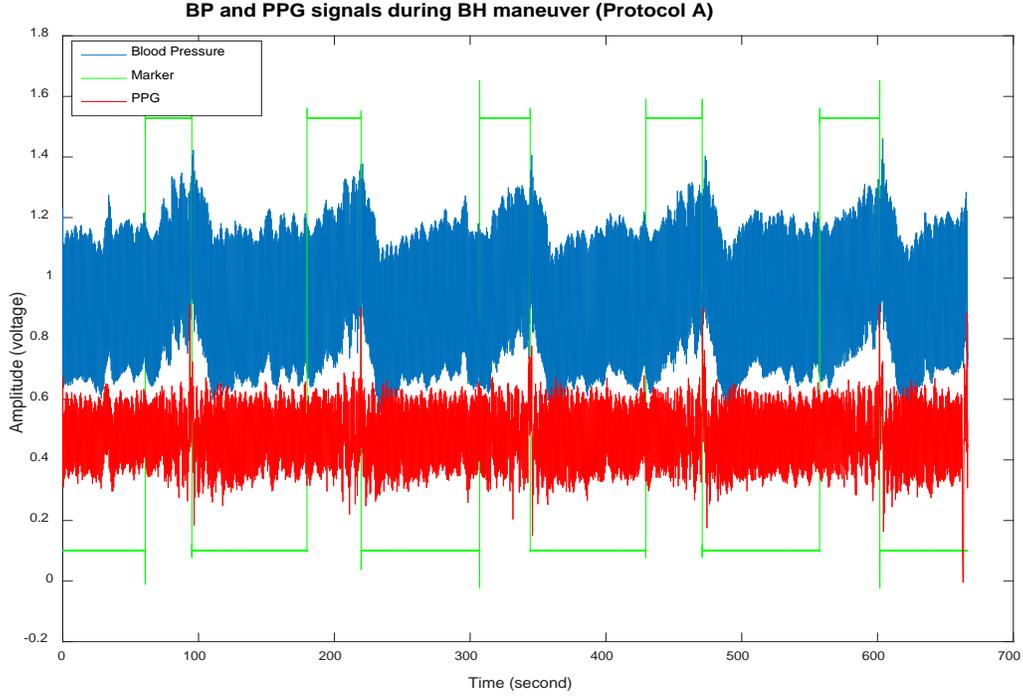

Figure 2- Recorded blood pressure (blue) and PPG (red) signals during BH maneuver. The green line shows the breath-hold maneuvers (BH, high values) and normal breathing (NB, low values) intervals.

*E. Model and prediction error*

To ascertain the accuracy of the models, once the model parameters (i.e. $a_i$'s, $b_j$'s, $n_a$, $n_b$, and $n_k$ in Eq. 1) for each experimental setting (i.e. BH1, BH2… BH5 and NB1, NB2… NB5) were determined, they were kept constant and used to compute the sample-by-sample difference between the model estimated BP features (i.e. $\widehat{SBP}$, $\widehat{DBP}$, and $\widehat{MAP}$) and the measured BP features values (i.e. SBP, DBP, and MAP, respectively). We refer to this difference as *model error*, as they reflect how well the model fits the experimental data for each experimental setting. Further, for each experimental setting, each BP feature, and for all subjects combined we aggregated the *model error* values by computing the root mean square of the errors using the following equation:

$$rMSE = \sqrt{\frac{\sum_{t=1}^{n}(\hat{y}_t - y_t)^2}{n}} \qquad (3)$$

Where, $y_t$ is the $t^{th}$ sample of the measured BP feature, $\hat{y}_t$ is the $t^{th}$ estimated value of the BP feature by the model, and n is the number of samples in the BP feature time series. Explicitly, when applying Eq. 3 the values of $\hat{y}_t$ and their corresponding $y_t$ values for each experimental setting, each BP feature, and all subjects combined were used to compute the rMSE values for modeling that feature. That is, under our experimental design, for each experimental setting and for each BP feature, there are 15 subject datasets of *model error* values that were used. Therefore, all *model error* values obtained from all subjects contributed to computation of the rMSE values for *model error*.

We also assessed the accuracy of each model that was derived from measured data for a subject and for a given experimental setting in predicting the measured values of BP features for other congruent experimental settings. For example, for each subject we applied the model that was derived from estimating SBP during BH1 to predict SBP for BH2, BH3… and BH5 and repeated the same for the model derived from BH2 data to predict SBP for BH1, BH3, BH4, and BH5 and so on. This resulted in 20 cases per each subject. With this method, the prediction accuracy was assessed for all combinations of the congruent experimental settings. We refer to the difference between each sample of predicted BP feature and its measured value as *prediction error*. To establish an aggregate representation of the *predication error* values, we applied Eq. 3 to all corresponding *prediction error* values obtained from all subjects. Hence, for each BP feature and each experimental setting, the *prediction error* values obtained for all subjects contributed to computation of the corresponding rMSE values of the *predication error*.

## III. RESULTS

Fig. 2 shows a sample of computed $\widehat{DBP}$, $\widehat{SBP}$, and $\widehat{MAP}$ superimposed on their respective measured values of DBP, SBP, and MAP. Fig. 3 shows the mean of the model errors as well as the mean of the prediction errors. Fig. 4 displays the standard deviation of the model errors and prediction errors reflecting the level of dispersion in these errors around their mean values.

To estimate the accuracy of the values of $\widehat{SBP}$, $\widehat{DBP}$, and $\widehat{MAP}$ obtained from each model, the root mean square error (rMSE) for the model errors for both BH and NB intervals were computed using (3). The results of averaging the rMSE values derived from the model errors for all subjects and for BH and NB intervals are tabulated in Tables I and II, respectively. To assess how a model developed from the data of one interval (e.g. BH1) predicts the corresponding BP measures (i.e. SBP, DBP, or MAP) of another congruent interval, we computed the rMSE of the prediction errors for NB and BH and tabulated them in Tables III and IV.

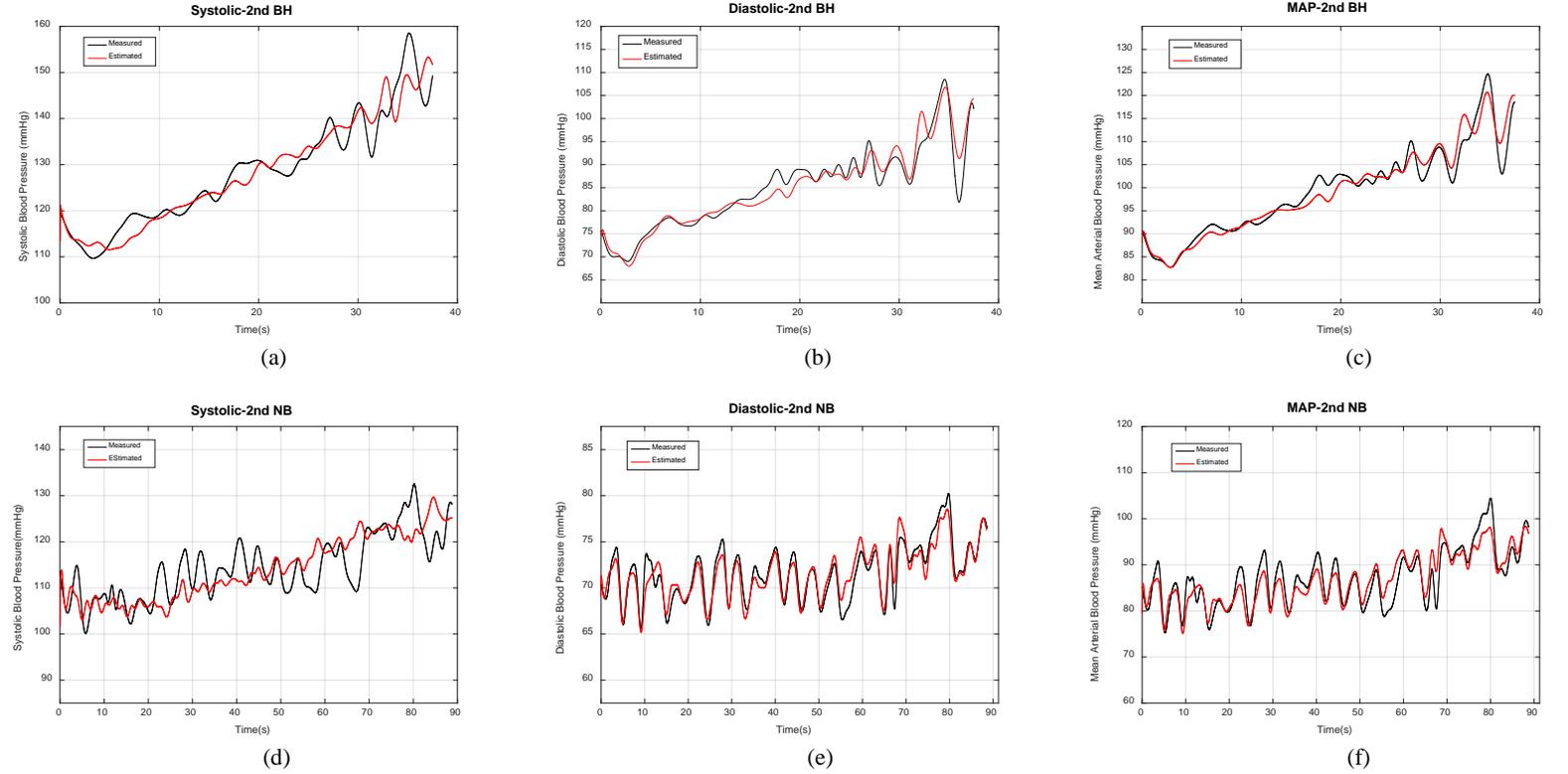

Figure 3- Measured BP signal versus estimated BP from modeling: figures (a), (b), and (c) show examples of breath-hold models and figures (d), (e), and (f) show examples of normal breathing models. All plots are for the same subject. Red plots are estimated signals and black plots are the measured signals.

Table I: rMSEs for models errors of NB intervals (mmHg)

|  | *1st NB* | *2nd NB* | *3rd NB* | *4th NB* | *5th NB* | *6th NB* |
|---|---|---|---|---|---|---|
| ***Systolic*** | 4.386 | 4.583 | 4.287 | 3.953 | 4.692 | 4.259 |
| ***Diastolic*** | 2.969 | 3.495 | 3.100 | 3.345 | 3.492 | 3.122 |
| ***MAP*** | 3.065 | 3.318 | 3.042 | 3.137 | 3.354 | 3.025 |

Table II: rMSEs for model errors of BH intervals (mmHg)

|  | *1st BH* | *2nd BH* | *3rd BH* | *4th BH* | *5th BH* |
|---|---|---|---|---|---|
| ***Systolic*** | 4.601 | 4.190 | 4.806 | 4.347 | 4.935 |
| ***Diastolic*** | 3.294 | 2.884 | 3.191 | 3.255 | 3.244 |
| ***MAP*** | 3.271 | 2.842 | 3.137 | 2.939 | 3.269 |

Table III: rMSE of prediction errors of models identified for each NP interval and applied to estimating BP features for other NB intervals (mmHg)

|  | 1st NB | 2nd NB | 3rd NB | 4th NB | 5th NB | 6th NB |
|---|---|---|---|---|---|---|
| *Systolic* | 6.582 | 8.137 | 6.681 | 7.875 | 7.944 | 6.755 |
| *Diastolic* | 4.013 | 6.246 | 6.033 | 6.439 | 5.889 | 6.092 |
| *MAP* | 4.224 | 5.736 | 5.269 | 5.847 | 5.640 | 5.322 |

Table IV: rMSE of prediction errors of models identified for each BH interval and applied to estimating BP features for other BH intervals (mmHg)

|  | 1st BH | 2nd BH | 3rd BH | 4th BH | 5th BH |
|---|---|---|---|---|---|
| *Systolic* | 7.375 | 6.630 | 6.481 | 6.493 | 7.944 |
| *Diastolic* | 4.478 | 4.239 | 4.629 | 4.334 | 4.652 |
| *MAP* | 4.655 | 4.358 | 4.459 | 4.225 | 4.405 |

To ascertain whether there was any difference among the models that were derived from the 15 subjects' data, we compared the means of model errors as well as prediction errors of models derived from each subject data with the corresponding errors for models derived from other subjects' data. Please note that since for each subject there were two respiratory modes (i.e. BH and NB), three blood pressure parameters of interest (i.e. SBP, DBP, and MAP), and two types of errors (i.e. model errors and prediction errors) then there were $2 \times 3 \times 2 = 12$ combinations that needed to be tested (i.e. number of cells in Table V). Further, since there were 15 subjects, 105 distinct comparisons of the errors for unique subject pairs needed to be performed. Using the analysis of variance (ANOVA) combined with multiple comparison, we tested for any significant difference ($p < 0.05$) between subjects' results [42]. Table V shows the number of unequal residual means that resulted from these comparisons. In the table, an entry of 4 in the cell at the intersection of the first column and first row indicates that only 4 out of 105 comparisons of model errors for SBP during BH intervals had unequal means. Considering that the totality of the comparisons of means of both the model errors and prediction errors equate to $12 \times 105 = 1260$ and the fact that a total of 9 mean comparisons tested to be unequal, the percentage of unequal means is 0.71%.

Table V: Number of statistically different means of errors among subjects ($p < 0.05$)

|  | BH | | NB | |
|---|---|---|---|---|
|  | Model | Prediction | Model | Prediction |
| *Systolic* | 4 | 3 | 0 | 0 |
| *Diastolic* | 0 | 0 | 0 | 0 |
| *MAP* | 1 | 1 | 0 | 0 |

## IV. DISCUSSION

In Fig. 3, the subplots (a), (b), and (c) show that the $\widehat{SBP}$, $\widehat{DBP}$, and $\widehat{MAP}$ track the overall rising trends of the DBP, SBP, and MAP signals during the BH intervals. As shown, there is an upward trend in BH intervals indicating that BP is increasing during BH intervals, which is a direct result of the sympathetic nervous system response to breath-hold maneuvers [43]. Further, these subplots illustrate the ability of the models to track the higher frequency modulations present in the measured BH signals. Similar observations can be made about the results plotted in the subplots (d), (e), and (f) of Fig. 3. As can be seen from these sample plots, the ARMA models developed can track the measured values of the BP parameters of interest for both modes of NB and BH with a relatively high level of fidelity. However, it can be seen that modeling results shown in Fig. 3d do not track all the oscillations of the measured SBP as closely as in the other subplots of Fig. 3. This is due to the fact that we have selected the maximum model orders for all experimental conditions to be 5, based on evaluation of a sample number of records. This was done to avoid custom fitting the data for each experimental setting and to achieve some degree of uniformity in the derived models. Hence, some instantaneous estimates of the BP features may follow the measured values more closely than the others, but, as shown in all subplots of Fig. 3, the estimates do follow the longer term trends of the measured values. Further, quantitative comparison of results, which are shown in Tables I and II, show all computed model errors were less than 5 mmHg. Some models may perform better than the others depending on the frequency content of that specific interval or baseline trend.

The plots of the means of the model errors and prediction errors in Fig. 4 provide a visual assessment of the level of accuracy of estimation of the SBP, DBP, and MAP. As can be seen from the subplots (a), (b), and (c) of Fig. 4, the estimation errors for NB interval in all cases had a mean that was well within ± 3 mmHg. Indeed, a majority of the cases had error means that were within ± 2 mmHg. With taking the level of dispersion of the model errors in Fig. 5 into consideration, it can be seen that the standard deviations for the modeling errors for BH intervals were all below 5 mmHg. This relatively small window of variation was also corroborated by the congruent rMSE values that are shown in Table I. A similar observation can be made regarding the errors for BH interval. As illustrated, the mean values of these errors fell within a tighter interval than those of NB; all fell within ± 2 mmHg. Moreover, the dispersion of the errors for BH was comparable to that of NB by being below 5 mmHg.

As can be seen from the results shown in Fig. 5, the upper boundary for the dispersion of the prediction errors was larger than the dispersion for model errors (i.e. 10 vs 5 mmHg, respectively). This was somewhat expected as prediction errors reflect the ability of the models to predict BP under conditions that differ from the conditions from which that model was derived. In particular, the experimental protocol was designed to examine the effect of successive breath-holds on the estimation of BP.

Tables I and II show additional information about the dispersion of the estimates of SBP, DBP, and MAP. The rMSE values for the model errors for all NB and BH cases laid below 5 mmHg. This indicates that the developed models from the measured data, on the average, had an accuracy of 5 mmHg when they were applied to predict the SBP, DBP, or MAP for the same interval.

The overall dispersion of the errors of a model in predicting values of SBP, DBP, and MAP by models derived from a given interval and applied to different, but congruent (i.e. BH to BH and NB to NB), intervals are shown in Tables III and IV. Considering that the model parameters were derived from a different interval, one can expect the estimation errors to rise. Indeed, comparing the rMSE values in Tables I and II with those in Tables III and IV shows that rMSE values have a max mean of approximately 8 mmHg. Therefore, if rMSE is used to gauge the level of the error, for both model errors and prediction errors, an overall error of less than 8 mmHg can be expected. When compared with some of the previously reported techniques, one finds that our results are comparable to techniques that used PAT in estimating BP from PPG signals [44]. The mean error reported in the previous study was 5 mmHg for DBP and 1 mmHg for SBP, while our method achieved mean error of less than 3 mmHG for all SBP, DBP, and MAP [45]. PTT has been used to estimate mean SBP and mean DBP and compared to actual measurements and reported to be within 15 and 25 mmHg, respectively, which is much higher than our method. These findings show the ARMA model approach is capable of tracking slow frequency trends and also high frequency hemodynamic changes of the body and exhibits adequate accuracy for possible clinical applications. Nonetheless, since the blood pressure level is influenced by a host of variables such as cardiac output and peripheral vascular resistance, introducing other measurable parameters, particularly from PPG, may increase the accuracy of blood pressure estimation [46].

The results shown in Table V illustrate that only less than 1 percent of model errors and predication errors had unequal means. This can be interpreted as that almost all errors for the models, regardless of the subject dataset that was used to obtain the model parameters, had the same mean value ($p < 0.05$). Therefore, this result suggests the selected model structure ((2) and max values of $n_a$, $n_b$, and $n_k$) has comparable accuracy for modeling BP for all the subjects in the sample population. It is noted that the few unequal error means reported in Table V (9 out of 1260 cases or 0.07%) occurred for BH intervals, and not during NB intervals. As expected, the changes in BP during BH intervals were higher due to the sympathetic nervous system response to breath-hold. Further, the intensity of these variations induced by breath-hold varied among subjects. In these few cases (i.e. 0.07%), the presence of unequal means for the errors may be attributable to these variations.

Results of this study suggest that with further development it may be possible to get reasonable estimates of key features of beat-to-beat blood pressure (i.e. SBP, DBP, and MAP) from PPG, creating opportunities for improvement in monitoring cardiovascular system health. One such opportunity is the ability to nocturnally measure blood pressure as part of now widely available and reasonably-priced pulse oximeter systems. Investigators have reported significant oscillations in nocturnal blood pressure due to obstructive sleep apnea [47]. However, due to cost and complexity, the current technology makes direct measurement of blood pressure during multiple nights at home impractical. A PPG-derived blood pressure measurement can provide for multiple night or long term measurement of blood pressure and might prove to be a valuable means for assessing the efficacy of therapy and/or medication for control of BP in sleep apnea patients.

It is important to note that the PPG signal reflects both blood oxygenation and volume changes; however, the main effect is dominated by volume change [41][47] [48]. As can be seen in Fig. 2, concurrent with changes in the BP waveform, the peak-to-peak amplitude and underlying frequency of the PPG signal also changed as the breath-hold maneuver progressed. The changes in the PPG may be partly attributable to the changes in the blood oxygen saturation, but they are mainly due to higher volume of the blood passed to the peripheral vasculature because of an increase in blood pressure.

In this study, we used two sets of models to achieve reasonable accuracy in predicting BP. If the presented method is to be applied for estimation of blood pressure due to some physiological perturbations such as obstructive sleep apnea, one would need to derive one model for normal breathing periods and another for apneic episodes. However, in case of sleep apnea, obtaining two such models is quite feasible, as apneic episodes are definitively identifiable through various means such as scoring of polysomnography data, respiratory monitoring, or analysis of oximetry data. Therefore, while needing two models requires added computations, it yields consistent accuracy in the estimation of BP (Tables I through IV), whether the patient is breathing normally or experiencing apnea.

While the aim of this study was to present a method of estimating BP from PPG, future development of the methods presented here will provide an opportunity for implementation in wearable health monitoring devices. Such implementation could bring about a positive impact for near-continuous monitoring of BP in a large sector of the population ranging from athletes to hypertensive patients. If such applications are considered, then steps need to be taken to eliminate or minimize the effect of any motion artifact [49] [50] [51]. Among many possible applications, such as the examples mentioned here, the methodology for a person-specific model calculation and validation needs to be developed. The likely need for person-specific models stems from the fact that there is a wide variation in the physiological systems involved in the control of blood pressure, which includes responsiveness of the sympathetic nervous system, mechanical, fluid mechanics, dynamical properties of the cardiovascular system, and metabolic rate. In this study, we explored only the linear ARMA models since they provide distinct advantages in analyzing the dynamic characteristics of the modeled system (e.g. stability, frequency response assessment, etc.) and the existence of a well-established body of knowledge about linear systems. Further development can involve expansion of this method to include the use of nonlinear modeling, as it may increase the accuracy of the results.

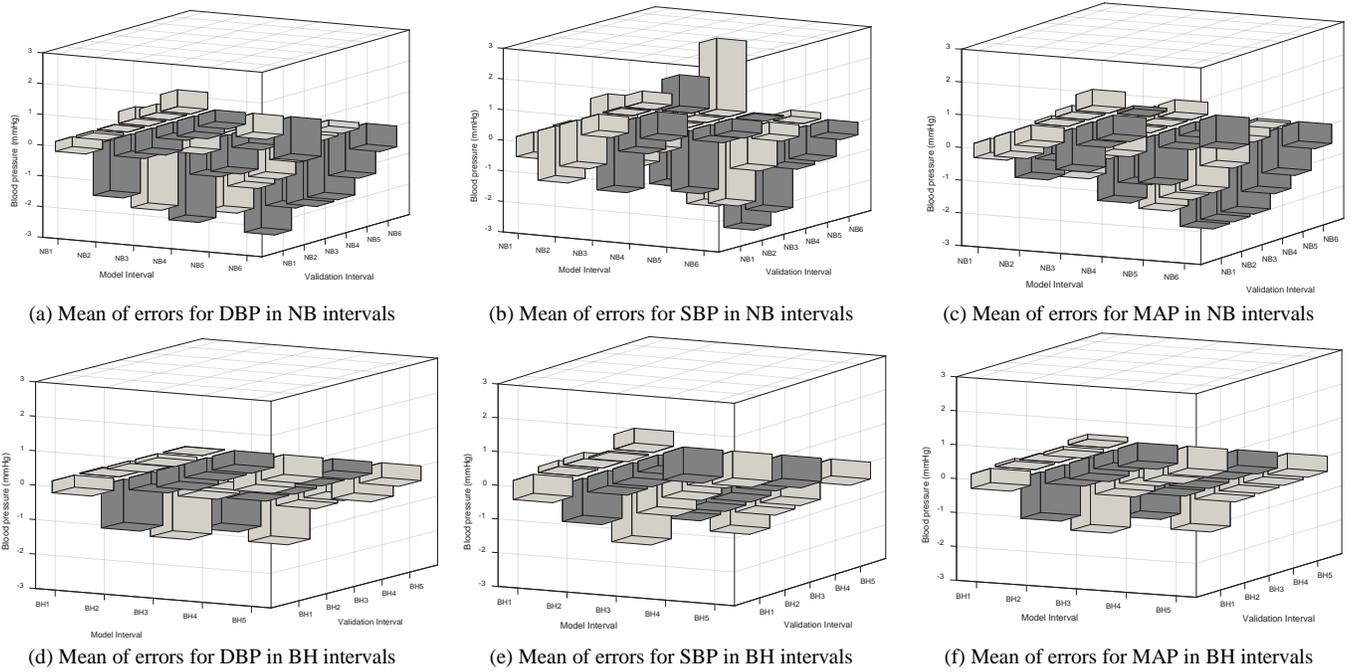

Figure 4- Mean of errors for DBP, SBP, and MAP in both NB and BH intervals. Each model was evaluated by all other congruent models.

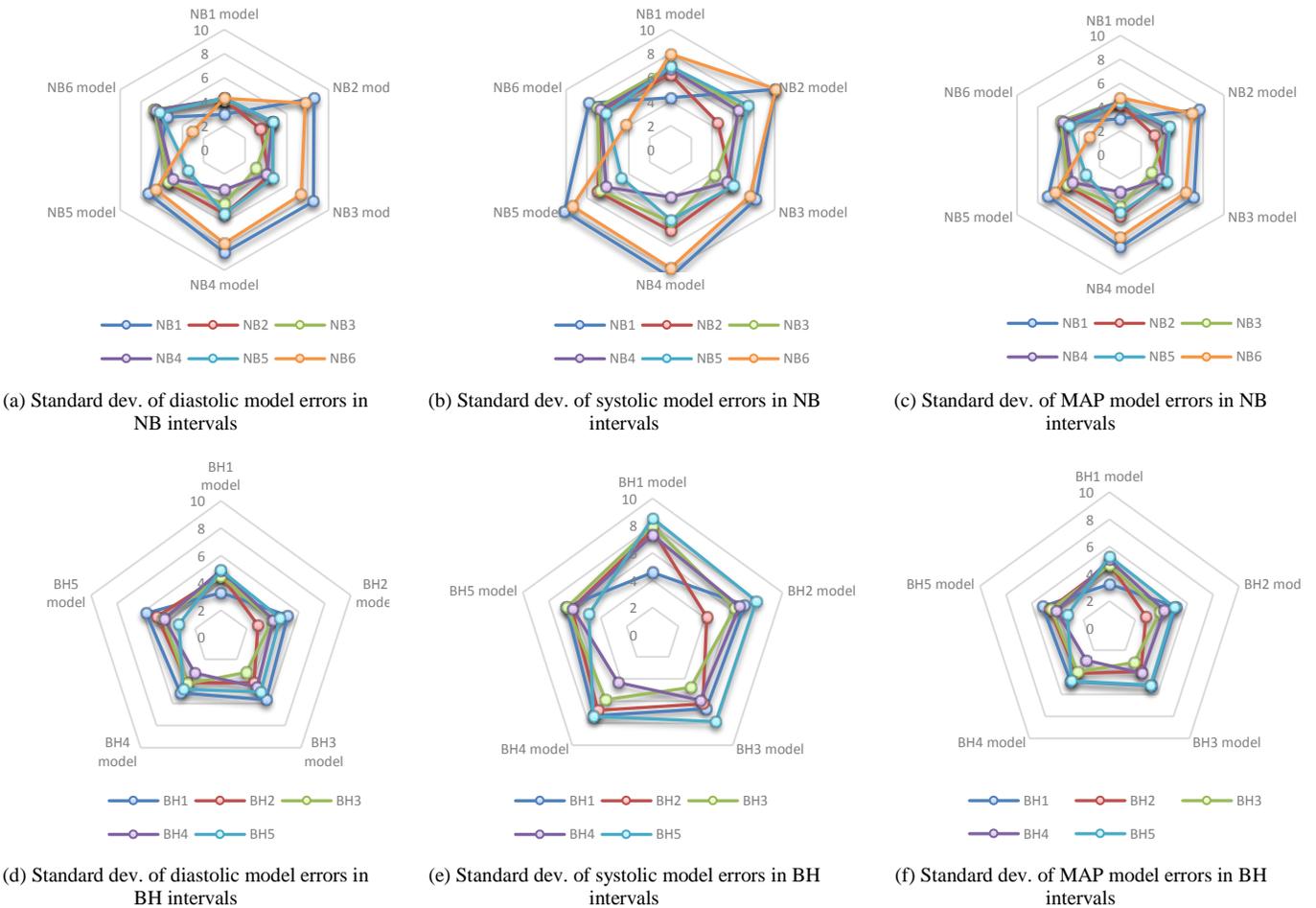

Figure 5- Standard deviation of errors for DBP, SBP, and MAP in both NB and BH intervals. Each model was evaluated by all other congruent models.

## V. Conclusion

The findings of this study demonstrate that estimating systolic and diastolic BP from PPG measurements using ARMA models can be a viable method for continuous and noninvasive measurement of key BP focal points with a high level of accuracy. Further, the low level of estimation error in both normal breathing and breath-hold maneuvers makes the proposed method a highly promising approach for certain clinical applications, such as longer term ambulatory and nocturnal blood pressure studies.